\documentclass[12pt,a4paper,twoside]{scrartcl}
\usepackage{graphicx}
\usepackage{amsmath}
\usepackage[figuresright]{rotating}
\usepackage{fancyhdr}
\usepackage[width=16cm]{geometry} 
\usepackage{subfigure}

\newcommand{\TheAuthor}{}

\newcommand{\GF} {\it G_{F}}
\newcommand{\gp} {\it g_{P}}
\newcommand{\lop}{\lambda_{OP}}
\def\skiplinehalf{\medskip\\}

\def\supit#1{\raisebox{0.8ex}{\small\it #1}\hspace{0.05em}}  
\title{
\begin{center}
New precision determination of $\gp$ and $G_{F}$: the MuXperiments at PSI
\end{center}}
\author{
Bernhard Lauss \supit{} 
\skiplinehalf
\supit{} {\normalsize Physics Department, University of California,} \\
{\normalsize 366 LeConte Hall, Berkeley, CA 94720, USA} \\
{\it \normalsize lauss@berkeley.edu} \\
\skiplinehalf
{\normalsize on behalf of the MuCAP~\cite{mucap} and MuLAN~\cite{mulan} collaborations}}
\date{}
\begin{document}
\maketitle

\vspace*{-15mm}

\begin{abstract}
We discuss two precision experiments which will measure fundamental weak interaction parameters:
MuLAN's goal is the measurement of the positive muon lifetime to 1~ppm, 
which will in turn determine the Fermi coupling constant $\GF$ to 0.5~ppm precision.
MuCAP is the first experiment which will unambiguously determine the induced pseudoscalar
form factor of the proton, $\gp$. While contradictory experimental results 
for $\gp$ are under discussion,
firm theoretical calculations on the percent level within the framework of Chiral Perturbation Theory
are now challenging the measurements. 
We will describe our experimental efforts and latest achievements.
\end{abstract}
%

\section{Precision determination of $\GF$ - the MuLAN experiment}

\begin{figure}[h]
\begin{center}
\includegraphics[height=40mm]{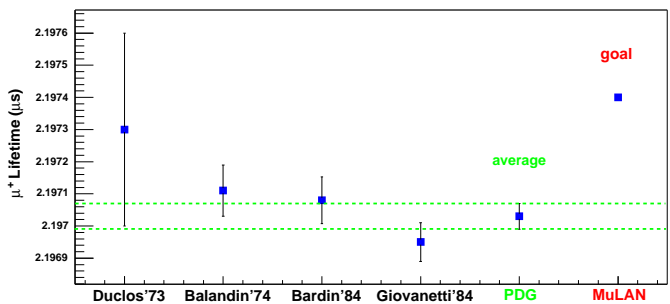}
\caption{Progression of positive muon lifetime measurements \cite{PDG}. 
The 1~ppm error goal of MuLAN is too small to be visible on this scale.
}
\label{mulife_history}
\end{center}
\end{figure}

We have seen impressive advances in our precise knowledge
of many parameters defining the electroweak interaction within the Standard Model.
However, the value of one of the most fundamental weak parameters, 
the Fermi coupling constant $\GF$, has not been improved in over
two decades (see Fig.\ref{mulife_history}).
Usually $\GF$ is determined via a measurement of the 
muon lifetime $\tau_\mu$
\begin{equation}
\it
\frac{1}{\tau_{\mu}} = \frac{G_F^2 m_\mu^5}{192 \pi^3} F\biggl(\frac{m_e^2}{m_\mu^2}\biggr) 
\biggl\{1+\frac{3}{5}\frac{m_\mu^2}{M_W^2}\biggr\}
\biggl(1 + \Delta_{QED}(\alpha_{m_\mu})\biggr)  ~,
\end{equation}
with
$F(x) = 1-8x-12x^2$ ln$x+8x^3-x^4$ \cite{musolf}.
The QED corrections within the Fermi Model, $\it \Delta_{QED}$, are included in this definition.

Within the Standard Model one can derive the relation 
\begin{equation}
\GF = \frac {\pi \alpha(0)}{\sqrt2 M_W^2 \left(1-\frac{M_W^2}{M_Z^2}\right)} (1+\Delta_r) ~,
\label{mwprediction}
\end{equation}
with weak radiative corrections being summarized in $\it \Delta_r$ \cite{hollik}.
The calculated quantity $\it \Delta_r$
depends on the entire set of input parameters, e.g. $M_Z$, $M_{Higgs}$, $m_{top}$, $\alpha$.
Recently, these calculations were improved to the sub-ppm level
by including numerically important
QCD and electroweak higher-order terms up to 2-loop level.
Therefore a precise comparison between theoretical and
experimental values, e.g. for $M_W$ (Eq.~\ref{mwprediction}) is possible \cite{hollik}.
Consequently, $\GF$ sets important constraints on the Standard Model 
and SUSY parameters. 
Furthermore, $\GF$ sets the weak scale and is intimately connected
to the vacuum expectation value of the Higgs field.
The best possible experimental measurement of $\GF$ 
at the present technological limit is therefore highly
desirable, as the 18~ppm precision limit on the PDG average on
$\tau_{\mu}$ \cite{PDG} is dominated by experimental counting statistics.

The MuLAN experiment ({\bf Mu}on {\bf L}ifetime {\bf AN}alysis), intents
to measure a total of $10^{12}$ $\mu^{+}$ decays, in order
to achieve a 1~ppm statistical error in the lifetime.
Since the status report on the MuXprogram in \cite{kammelexa02} we have achieved substantial progress.
A modification of the continuous high intensity muon beam line 
at the Paul Scherrer Institut was necessary 
to enable the collection of
$10^{12}$ events within a reasonable time.
We have built an electrostatic kicker 
which applies an artificial time structure to the 
DC beam in the $\pi$E3 area
and found a kickable beam tune which provides up to 8~MHz of muons.
Following a 5~$\mu$s muon collection period in the target, the 
kicker deflects the beam for 27~$\mu$s while muon decays are measured.

\begin{figure}[ht!]
\begin{center}
 \subfigure[]
{\includegraphics[height=60mm,width=.48\textwidth,angle=0]{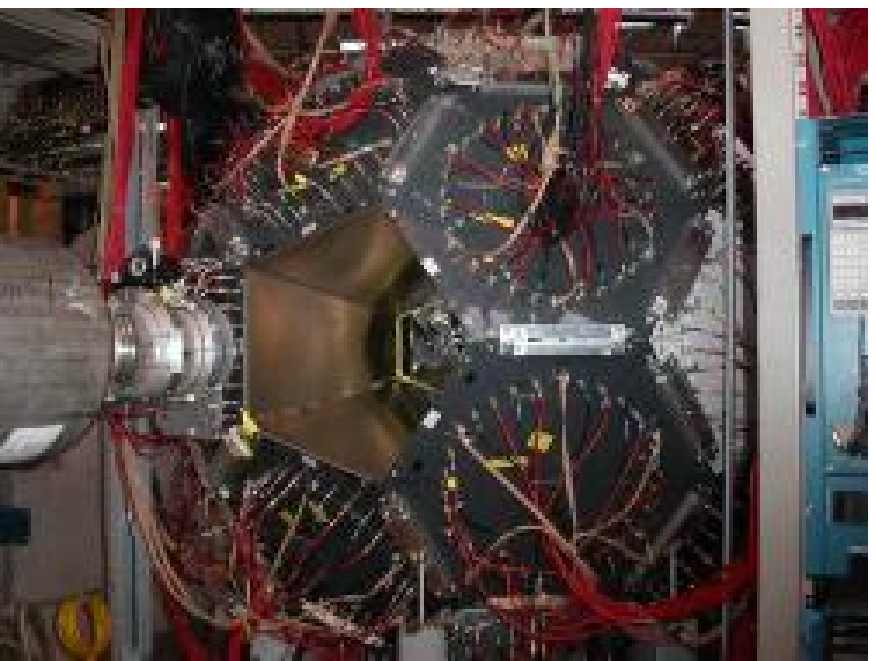}}
\subfigure[]
{\includegraphics[height=60mm,width=.41\textwidth,angle=0]{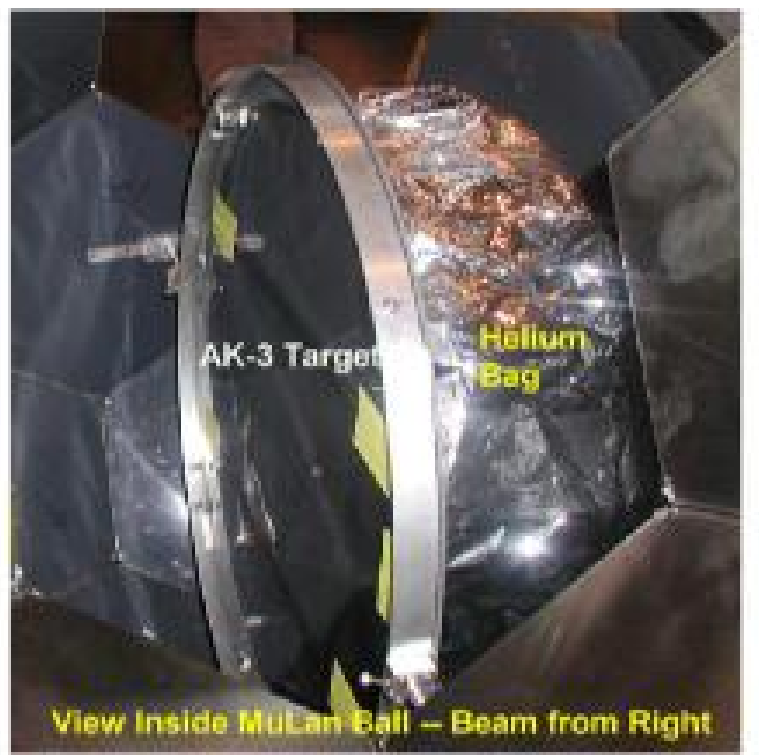}} 
\vspace*{-3mm}
\caption{The installed MuLAN setup:\newline
a) View of the soccer ball detector. The visible hexagonal structures are made of six 
neighboring triangular scintillator pairs.\newline
b) The Helium bag and the AK3 target mounted in 
the detector center.}
\label{mulan-setup}
\end{center}
\end{figure}

MuLAN is designed to minimize the systematic errors in several ways:\newline
{\bf $\bullet$ Muon polarization:} The beam muons are highly polarized, and the
preferential emission of decay positrons in muon spin direction 
could cause a position- and time-dependent positron detection efficiency
as polarized muons rotate in an external magnetic field. 
We are currently investigating two specific targets: 
i) Arnokrome-3 (Fig.\ref{mulan-setup}b) is a proprietary chromium-cobalt-iron alloy sheet, which, 
due to an internal field of a few Tesla,
precesses muons very fast with respect to muon decay.
Therefore polarization effects are negligible.
ii) A solid sulfur target which maximizes the
depolarization of the beam muons.
It is placed in a homogeneous 120~Gauss magnetic field
which causes a fast visible muon rotation and allows us to fit the
corresponding decay positron asymmetry.
We are presently investigating samples of $10^{10}$ decay positrons
from each target to select the optimal material choice.
Additionally, a polarization-preserving silver target is being used for
control purposes.\newline
MuLAN's highly modular detector (Fig.\ref{mulan-setup}a) of 174 coincident 
scintillator tile pairs in ``soccer ball'' geometry
allows us to compare opposite counters, thus strongly reducing precession 
effects in the count rate sum.

{\bf $\bullet$ ``Sneaky muons:''} A fast thin entrance muon counter (EMC) records beam muons and looks for
muons sneaking in during the measurement interval.
A magnet positioned behind the EMC precesses the tiny
fraction of muons stopped in the detector materials,
otherwise they too could cause small detection inefficiencies.

{\bf $\bullet$ Off-target muon stopping:} 
Muon stops before the target are minimized by
decreasing the materials in the muon path. Consequently
we installed a helium bag (Fig.\ref{mulan-setup}b) and used
very thin mylar windows and EMC materials.

{\bf $\bullet$ Pile-up:} 
The high detector modularity and the fast scintillator response time
reduce pile-up.
Additional time resolution will be gained 
via new 500~MHz waveform digitizers (WFD) presently under construction,
which will provide a double pulse resolution better than 4~ns. 
Final WFD implementation to all detector channels is planned 
for 2005.

\begin{figure}[h]
\begin{center}
\includegraphics[height=6cm]{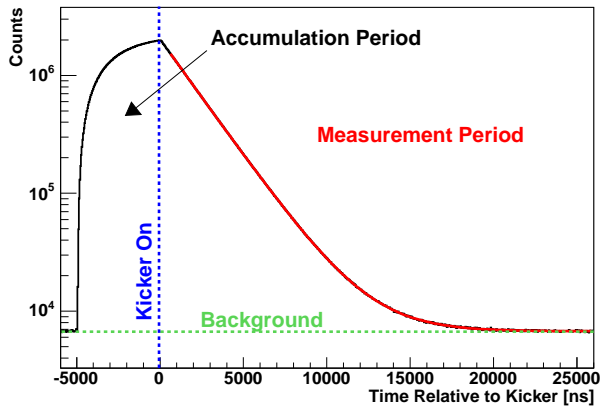}
\caption{
Lifetime spectrum (decay positron minus start of beam kick time)
obtained in a single 10~minute run displaying
accumulation and measurement period.  
}
\label{mulan-lifetime}
\end{center}
\end{figure}

The MuLAN detector (Fig.\ref{mulan-setup}a) was successfully commissioned in 2004 
and yielded its first physics data.
Fig.\ref{mulan-lifetime} shows a 10 minute snapshot from our AK3 target data.
We used multi-hit TDCs for detector readout.
The muon accumulation time and the decay recording time are indicated.
Our present analysis goal with this data is 
a 5~ppm precision determination of $\GF$. 
We intend to collect the full statistics in 2006.


\section{Precision determination of $\gp$ - the MuCAP experiment}

The $V-A$ description of weak interactions has been tested to a high precision.
Processes involving structureless fermions, e.g. muon decay, show equal 
vector ({\it V}) and axial-vector ({\it A}) coupling.
In $\beta$-decay as well as in 
nuclear muon capture on the proton \newline
$ \mu^{-} + p \rightarrow \nu_\mu + n$ ~~,
the axial coupling $G_A$ is modified due to hadronic effects
caused by the involved nucleon.
Muon capture occurs at higher four-momentum transfer $q=-0.88m_\mu^2$ than $\beta$-decay.
Lorentz invariance constrains the corresponding weak current matrix elements to six independent terms,
\begin{eqnarray}
   V_\mu & = &  G_V(q^2) \gamma_\mu - \frac{i G_M(q^2)}{2 m_N} \sigma_{\mu\nu} q^\nu + \frac {G_S(q^2)}{m_\mu} q_\mu \\
   A_\mu & = &  G_A(q^2) \gamma_\mu \gamma_5 + \frac{G_P(q^2)}{m_\mu} \gamma_5 q_\mu + \frac{i G_T(q^2)}{2 m_N}\sigma_{\mu\nu} q^\nu \gamma_5  ~,
\end{eqnarray}
with corresponding weak form factors $G_i$ ($_i$ = scalar, pseudoscalar, vector, axial-vector, tensor, weak magnetism);
mass of the nucleon $m_N$ and muon $m_\mu$.
Because of G-symmetry $G_S$ and $G_T$ vanish \cite{weinberg}.
Due to the momentum dependence, only $G_A$ and $G_V$ contribute in $\beta$-decay at very low $q^2$.
Nuclear muon capture is the process most sensitive to $G_P$.
Therefore, $G_P(-0.88m_\mu^2)$ is dubbed induced pseudoscalar coupling constant $\gp$. 
While the values of $G_V$, $G_A$ and $G_M$ are established 
on the $10^{-3}$ to $10^{-4}$ level \cite{PDG}, 
the situation is totally different for
the induced pseudoscalar $\gp$.

The theoretical view, historically based on PCAC and pion pole dominance,
and recently strictly derived within chiral perturbation
theory ($\chi$PT) \cite{meissner}, is remarkably precise:
\begin{eqnarray}
\label{gpequation}
 G_P(q^2) &=&  \frac{2 m_\mu g_{\pi NN} F_\pi}{m_\pi^2 - q^2} - \frac{1}{3 c^2 \hbar^2} G_A(0) m_\mu m_N r_A^2 ~, \\
 \gp  &=& (8.74 \pm 0.23) - (0.48 \pm 0.02) = 8.23 \pm 0.23 ~,
\end{eqnarray}
depending on the exact values of the pion-nucleon coupling constant $g_{\pi NN}$ 
and the mean axial radius of the nucleon $r_A$.
The Standard Model based calculation of  
the singlet muon capture rate by Govaerts and Lucio-Martinez \cite{govaerts}
has reached 0.55\% precision.
This precision in calculation will allow a high precision measurement
to distinguish between the pion pole contribution to $\gp$ and the 
correction term. Moreover, such a measurement will also set tight
limits on various theoretical scenarios 
beyond the Standard Model \cite{govaerts}.

\begin{figure}[t]
\begin{center}
\subfigure[]
{\includegraphics[angle=-90, width=.52\textwidth]{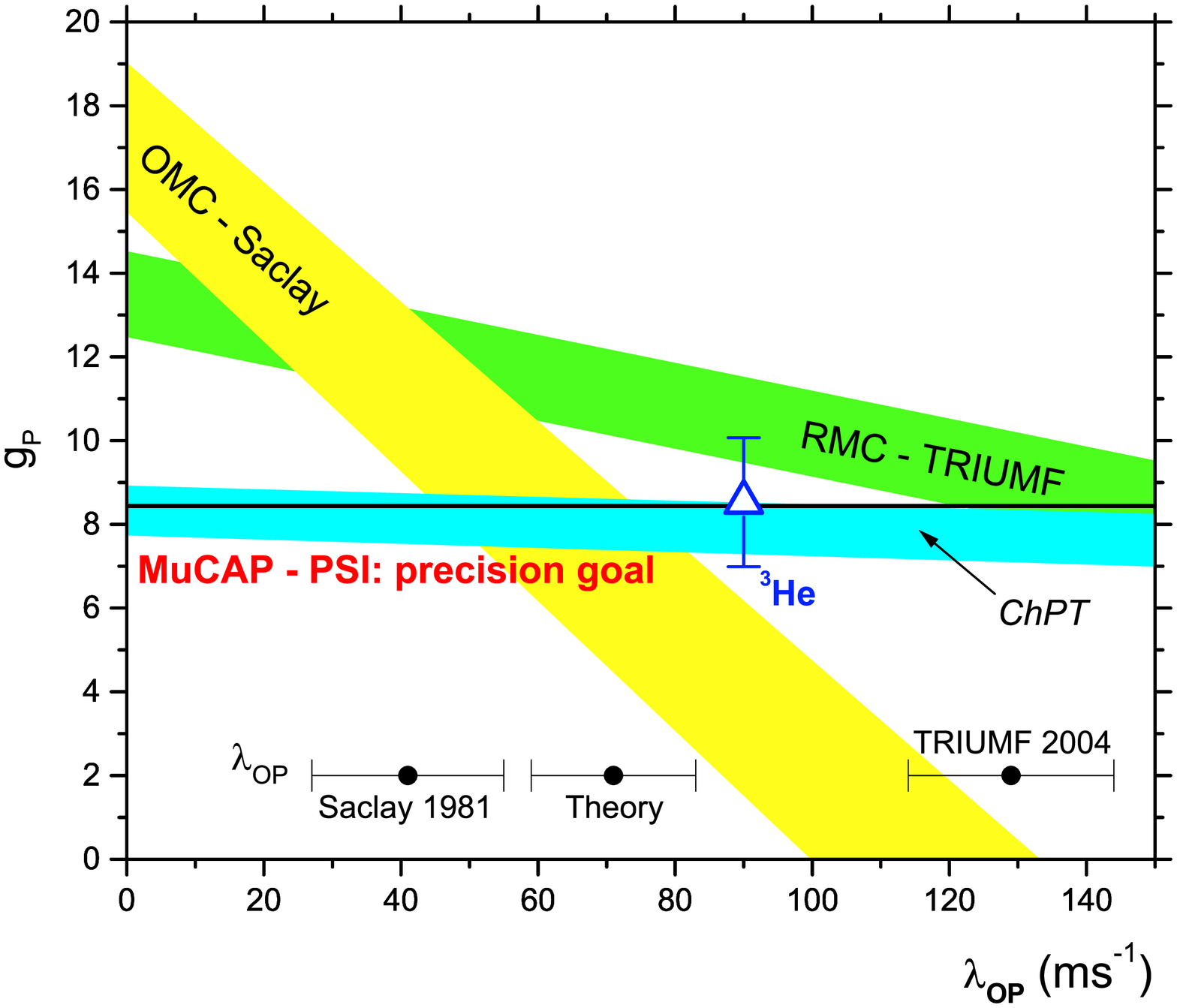}}
\subfigure[]
{\includegraphics[angle=-90, width=.47\textwidth]{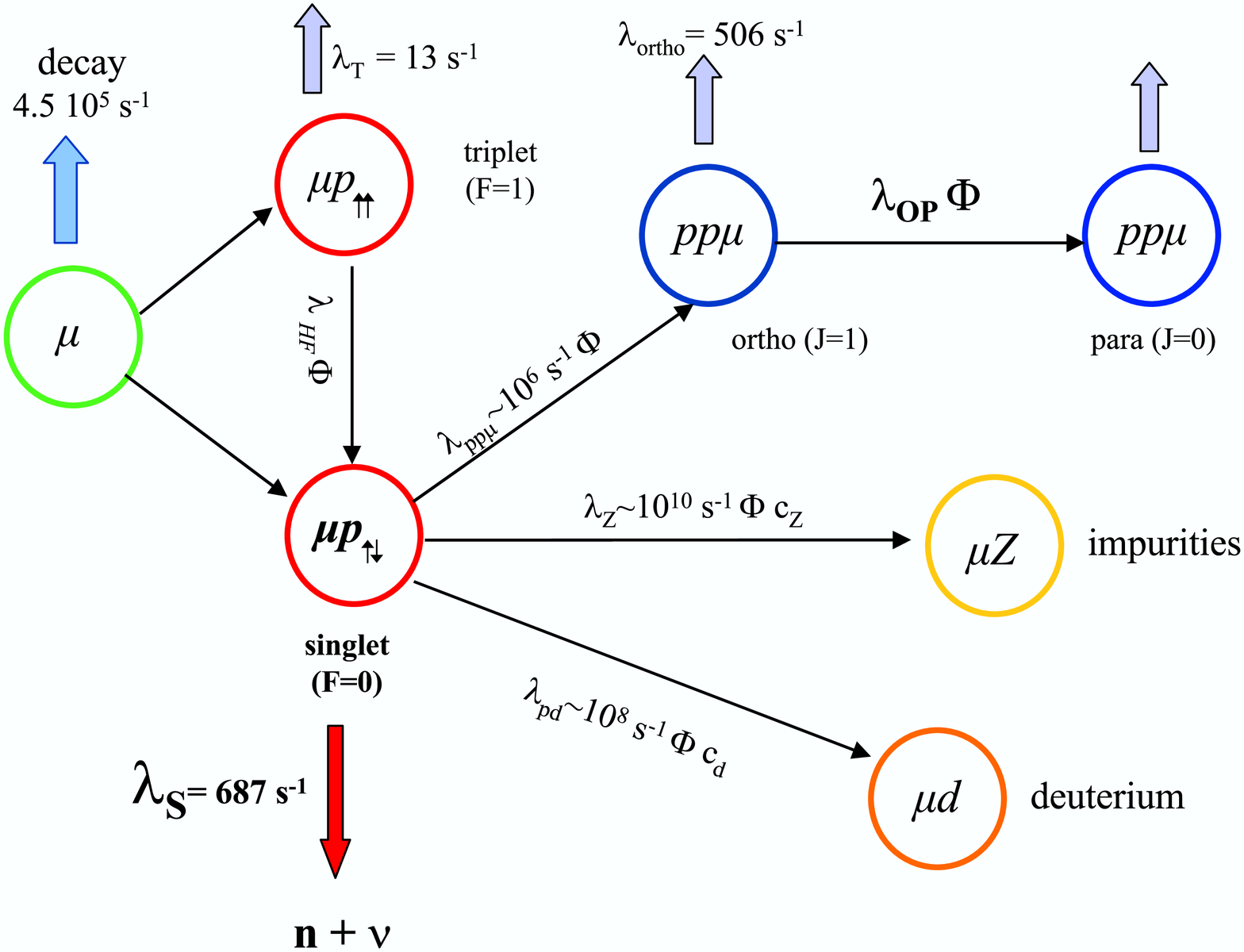}}
\vspace*{-5mm}
\caption{
a) Present knowledge of $\gp$:
The OMC \cite{saclay} and RMC \cite{rmc} results and their
$\lop$ dependence are shown 
together with the error goal for MuCAP plotted at the PCAC value.
The blue triangle shows the $^3$He result \cite{Ackerbauer}
at an arbitrary $\lop$ value.
The line shows the calculation from Ref.\cite{meissner}.
The $\lop$ values from experiments \cite{saclay-lop,triumf-lop} and theory 
\cite{bakalov} are indicated on the bottom.\newline
b) Schematic view of important muonic processes in a hydrogen
target, highlighting the capture in the $\mu p_{1s}$ state and 
competing background processes \newline
($\lambda$ = rate, $\Phi$ = density, c = concentration).
}
\label{gp-plot}
\end{center}
\end{figure}

The present experimental knowledge of $\gp$ is unsatisfying, 
and discrepancies cause considerable debate. 
Determinations via ordinary muon capture in hydrogen (OMC) \cite{saclay},
$^3$He \cite{Ackerbauer} and larger nuclei essentially confirm the theory result. 
However the precision of the latter is troubled by model dependencies. 
A radiative muon capture on the proton (RMC) experiment \cite{rmc}, 
which measured an additionally very rarely 
emitted high energy $\gamma$ ray 
in conjunction with the muon capture process, 
yielded a different result\footnote{While the RMC process 
has a 10$^5$ times lower branching than OMC, 
the emitted $\gamma$ can have energies up to 100~MeV. Therefore 
these $\gamma$'s come closer to the pion pole and the measurement is 
in principle four times more sensitive to $\gp$ than OMC.}.
The present most likely explanation
lies in the insufficient knowledge of the complex kinetics 
of negative muons in hydrogen.
The $\mu p$ atom is formed in statistically populated singlet and triplet states, and
the muon capture rate from these states differ by a factor of $\sim$60
due to the strong spin dependence of weak interactions. An exact
knowledge of the spin populations is therefore mandatory for the
interpretation of a measurement.
The initial $\mu p$ populations are modified via
spin flip and molecular formation processes, which eventually yield
a fraction of muonic hydrogen molecules ($pp\mu$). 
The respective rates for both types of process scale with density, and hence
are high in the liquid hydrogen targets which were used in Refs.~\cite{saclay} and \cite{rmc}.
However, a large $\mu$-molecular population 
causes a large correction to the lifetime and hence a corresponding uncertainty.
Specifically the ortho to para spin-flip rate in $pp\mu$ molecules $\lop$
(Fig.\ref{gp-plot}b) is a prime suspect to cause the experimental discrepancy in $\gp$.
Figure \ref{gp-plot}, updated from \cite{gorringe}, 
shows the $\lop$ dependence of the OMC and RMC results.
The controversial $\lop$ values are also shown. Two experimental values from 
Saclay \cite{saclay-lop} and TRIUMF \cite{triumf-lop}, 
which were obtained together within the same experiments 
performing the OMC and RMC measurements, strongly disagree, 
and the only theoretical calculation 
\cite{bakalov} does not clarify the situation.
It is evident that using the $\lop$ rate from the same measurements brings
the $\gp$ value in agreement with theory. 
However, the use of the other $\lop$ value enlarges the disagreement to the
prediction or lowers it to an unphysical negative value.
Clearly only a new determination of $\gp$
independent of $\lop$ can resolve this situation.


\begin{figure}[t]
\begin{center}
 \subfigure[]
{\includegraphics[height=60mm,width=.49\textwidth,angle=0]{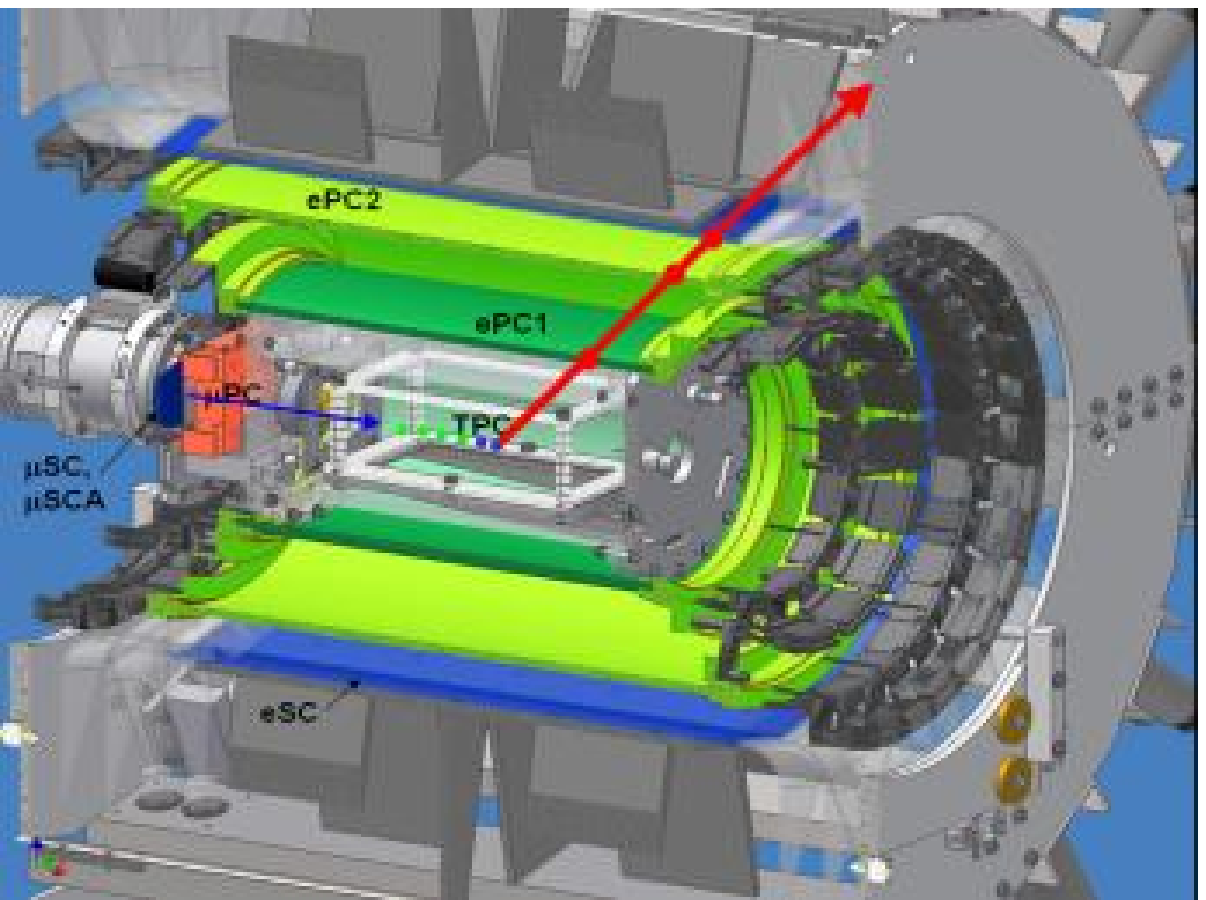}}
\subfigure[]
{\includegraphics[height=60mm,width=.49\textwidth,angle=0]{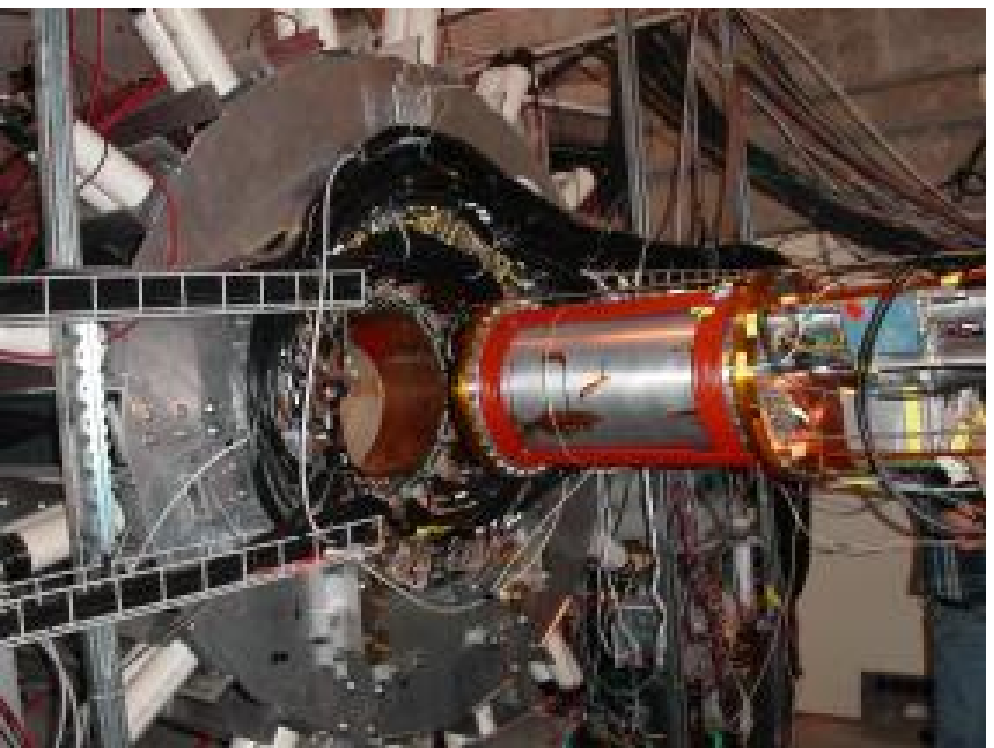}} \hfill
\vspace*{-5mm}
\caption{The setup of the MuCAP experiment:\newline
a) Schematic view of the detectors showing the
muon entering from the beam channel (blue arrow), 
hitting the 200 $\mu$m thick time=0 scintillator (muSC)
and a small wire chamber (muPC) which serves as additional pile-up counter,
and then stopping in the TPC.
The muon stop position is detected via 
a high signal caused by the Bragg peak (blue) after a track of low signals (green).
The decay electron (red) is observed in 2 cylindrical 
wire chambers (ePC1/2) for track reconstruction.
Outside is a doubled-layered 
scintillator hodoscope (eSC) with readout at both paddle ends, which records the decay time.\newline
b) Installed setup in PSI's $\pi$E5 area with the hydrogen vessel rolled back,
displaying the mounted $\mu$SR saddle-coil magnet.
}
\label{mucap-setup}
\end{center}
\end{figure}

\begin{figure}[t]
\begin{center}
\subfigure[]
{\includegraphics[width=.62\textwidth]{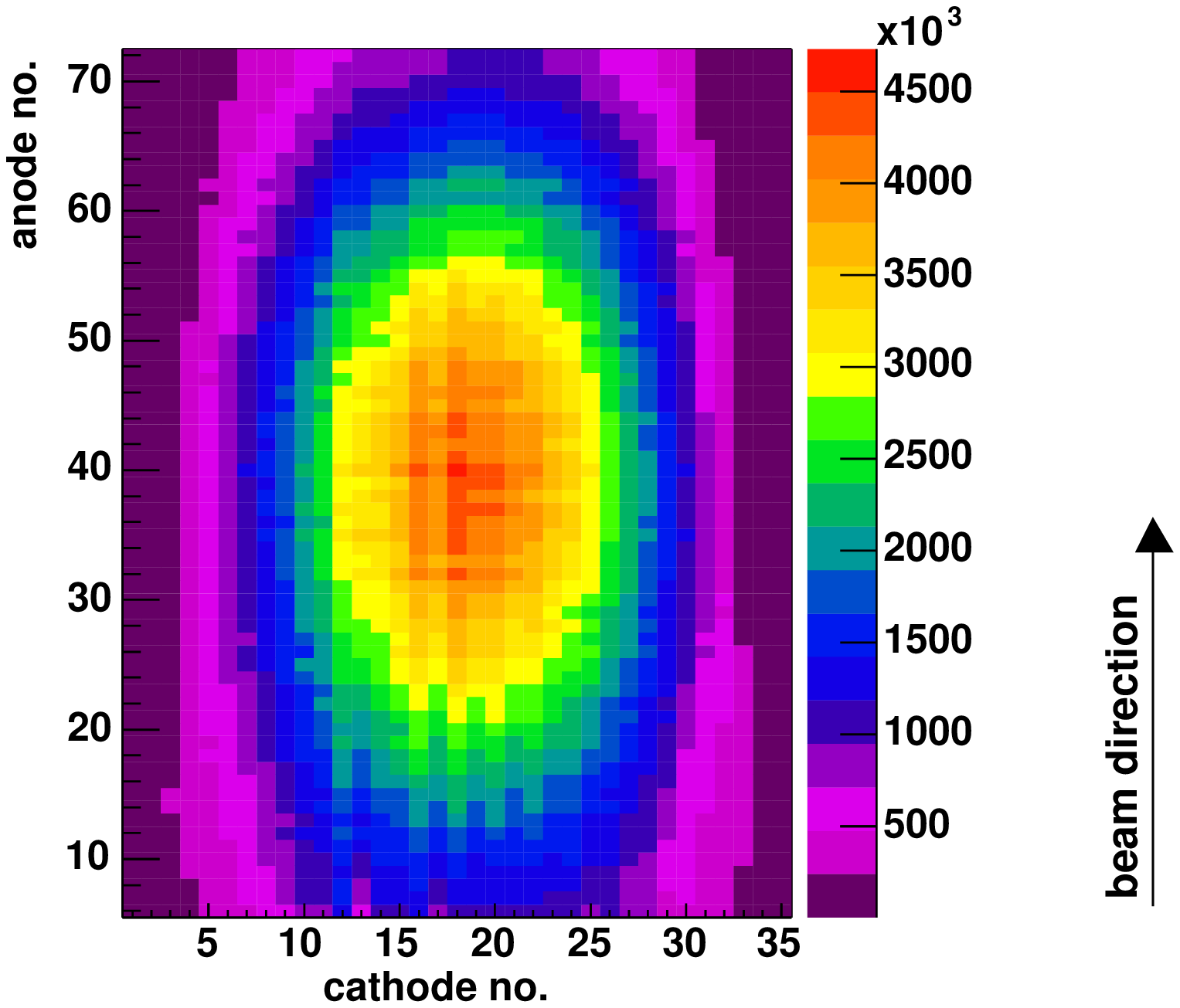}}
\subfigure[]
{\includegraphics[width=.36\textwidth]{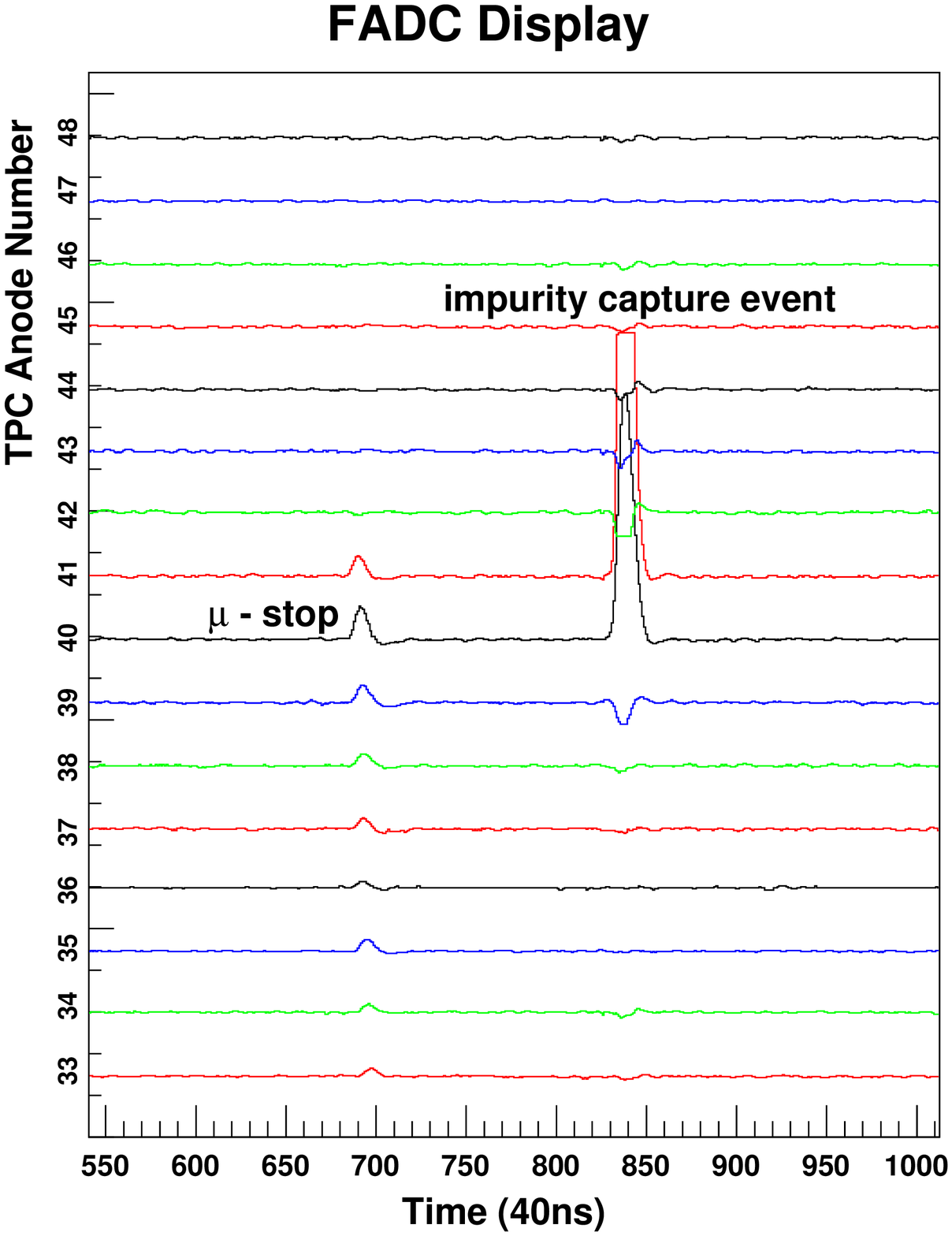}}
\vspace*{-5mm}
\caption{
a) Observed 2$\times$10$^9$ muon stopping positions in the TPC (fall 2004).\newline
b) Observed special event in the TPC: 
A muon stops after leaving a track over several anodes with a few times higher 
energy deposition in the last anodes. 
In fall 2004
we observed a fraction of $\sim$10$^{-6}$ high-$Z$ capture events, corresponding
to a 10$^{-8}$ concentration of impurities,
with a clear signature of 
a very high signal within 25~$\mu s$ after a muon stop.
}
\label{fadc_event}
\end{center}
\end{figure}

The experimental principle of the MuCAP ({\bf Mu}on {\bf CAP}ture) 
experiment is based on the measurement and comparison 
of the decay time of positive and negative muons in hydrogen.
The MuCAP experiment is designed to 
overcome the multiple difficult problems of previous experiments.
The important conceptual advantage of MuCAP is the selection of 
target hydrogen at gaseous density (10 bar at room temperature),
which minimizes the kinetics dependence of the result on $\gp$
as shown in Fig.\ref{gp-plot}a.
At low densities, muon capture occurs almost exclusively from the 
singlet state on the proton, and even a large estimated error on 
$\lop$ results only in a systematic error on the 10~ppm level.
The full setup is shown in Fig.\ref{mucap-setup}.
The active gaseous hydrogen target, a time projection chamber (TPC),
allows for a full 3-dimensional reconstruction of the muon path to its
stopping point and therefore a selection of clean muon stops
away from walls and wires. The observed stopping distribution of muons
is shown in Fig.\ref{fadc_event}a.

The TPC also detects muon capture events on 
impurity atoms ($Z > 1$) via the very large
signals generated from capture products.
Thus the TPC serves also as a very sensitive impurity monitor.
The high rates of muon transfer to and nuclear capture on high-$Z$ atoms 
(Fig.\ref{gp-plot}b)
can cause a deflection of the exponential lifetime even at very low impurity concentrations
as these rates are typically orders of magnitude higher than muon decay.
In order to minimize this effect, target purity requirements are very stringent, 
with the goal to be on the 10$^{-9}$ contamination level for the sum of high-$Z$ atoms.
Additionally, the exact knowledge of these contaminations is necessary to calculate the 
correction factor to the lifetime.
Consequently, the hydrogen gas 
after production in electrolysis is filled
via a palladium filter and continuously run through
a Zeolite based purification system (CHUPS). The CHUPS system \cite{chups} 
is specifically designed to maintain the hydrogen flow
with negligible variations in density or hence in TPC gain.
In fall 2004 we maintained 
clean target conditions (as low as 70~ppb impurities) for over 5 weeks.

\begin{figure}[t]
\begin{center}
\includegraphics[height=85mm,width=.99\textwidth]{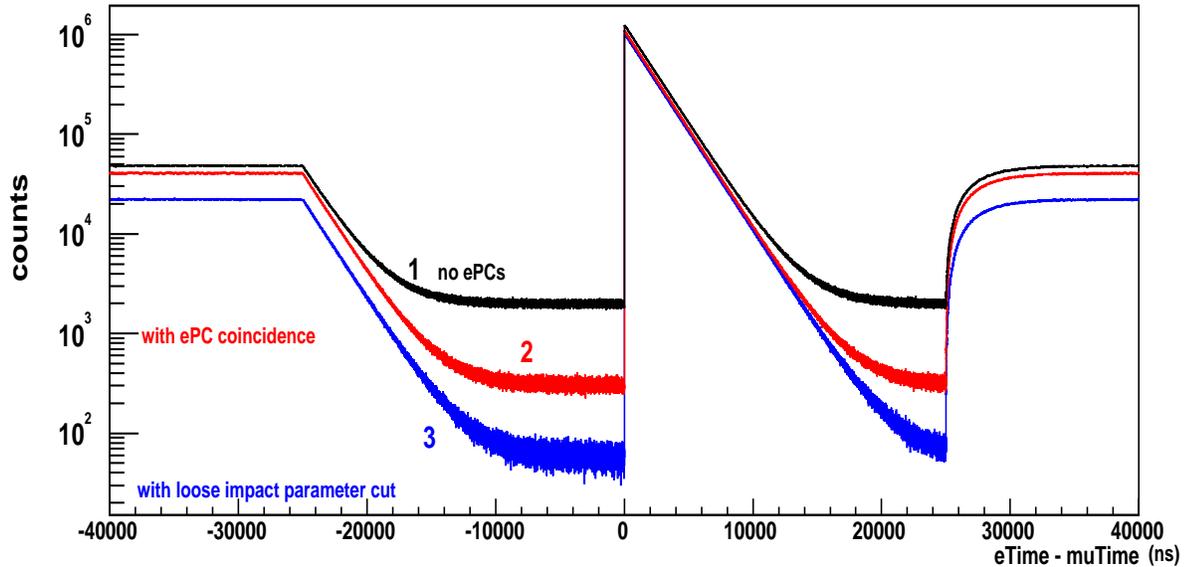}
\caption{
Preliminary muon lifetime spectrum
obtained with $\mu^-$
from the MuCAP fall 2004 run.
The three different curves show the benefits of the
decay electron detection by the two wire chambers.
Curve 1 (black) is obtained with requirement of a clean pileup-free muon stop
in the TPC and a four-fold eSC coincidence hit. Curve 2 (red) requires a 
time coincident hit in both ePC1 and ePC2.
Curve 3 (blue) additionally tracks the observed electron 
and requires less than 10~cm impact parameter
(approximately the sensitive TPC volume).
The huge reduction in background is obvious.
The necessary $\pm$25~$\mu$s pileup veto is responsible for 
the background's shape.
}
\label{mucap-lifetime}
\end{center}
\end{figure}

MuCAP fully separates the muon and electron detectors
to avoid dangerous cross-correlations.
Decay electron times are measured in the scintillator hodoscope (eSC) surrounding
the hydrogen vessel. 
Two cylindrical wire chambers track the electron back in 3 dimensions to its $\mu$-stop origin, 
thus largely reducing the background.

The impact parameter, defined as the minimal distance between detected muon stop and 
electron track, serves as an important handle on a very subtle systematic effect on the lifetime:
the diffusion of muons which transfer to deuterium, an isotope always present in hydrogen. 
Although we are using special deuterium-depleted hydrogen, (``protium''), 
with deuterium content as low as 1.5~ppm,
this deuterium concentration is still high enough to cause a visible
effect in our setup via electron tracks with dislocated origin.
This dislocation, due to $\mu d$ diffusion over macroscopic distances is possible because of
a Ramsauer-Townsend minimum in the $\mu d + p$ scattering cross-section at low energies.
Such $\mu d$'s can hit a wall or wire, transfer and then undergo unobserved 
nuclear capture outside the sensitive volume. 
The electron wire chamber tracking identifies such events
and the corresponding lifetime dependence on the impact parameter will allow us to determine
the deuterium contamination in situ \cite{kammel-mudcut}.
Additionally, we have been developing a precision trace deuterium
detection method via $pd\mu$ fusion \cite{pdmu-note}.

As in MuLAN, MuCAP applies a magnetic field
to control muon spin rotation effects in the $\mu^+$ measurement. A water-cooled aluminum 
coil was developed to minimize decay electron scattering.

Eventually MuCAP needs 10$^{10}$ cleanly observed decay electrons and positrons to
statistically reach the goal of 1~\% in the capture rate, reflecting 10~ppm in the 
respective muon lifetimes.
This is possible in several weeks of running in the fully pile-up protected mode.
In the future we also intend to use the MuLAN kicked beamline with the MuCAP experiment
and operate it in a ``muon on request mode'' \cite{kammel-more}.

Fig.\ref{mucap-lifetime} shows a preliminary lifetime plot from our fall 2004 data.
One can clearly see the huge improvement in background reduction due to the implementation
of the two cylindrical wire chambers.
The shown $\sim$2$\times$10$^9$ ``clean'' $\mu$ decay events are presently being analyzed,
and a first result on $\gp$ is expected in late 2005.

\section{Summary}

Both experiments, MuCAP and MuLan, are on the way to reach ultimate precision
in their respective measurements. First physics results are expected 
in late 2005.


%
\paragraph{Acknowledgments}
I would like to express my cordial thanks to all colleagues in the
MuLAN and MuCAP collaboration for creating a great scientific 
working environment and for their personal friendship.
This work was supported by the US Department of Energy and the
National Science Foundation.

\end{document}